# RESONANT AND NON-RESONANT TUNNELING THROUGH A DOUBLE BARRIER

S. V.Olkhovsky,* E.Recami,** A.K.Zaichenko*

(*) *Institute for Nuclear Research, NANU, Kiev-28, Ukraine,*

olkhovsk@kinr.kiev.ua

(**) *Facoltà di Ingegneria, Università statale di Bergamo, Bergamo, Italy,*

and *INFN-Sezione di Milano, Milan. Italy.*

recami@mi.infn.it

Abstract: An explicit expression is obtained for the phase-time corresponding to tunneling of a (non-relativistic) particle through two rectangular barriers, both in the case of resonant and in the case of non-resonant tunneling. It is shown that the behavior of the transmission coefficient and of the tunneling phase-time near a resonance is given by expressions with "Breit-Wigner type" denominators. By contrast, it is shown that, when the tunneling probability is low (but not negligible), the *non-resonant* tunneling time depends on the barrier width and on the distance between the barriers only in a very *weak* (exponentially decreasing) way: This can imply in various cases, as well-known, the highly Superluminal tunneling associated with the so-called "generalized Hartman Effect"; but we are now able to improve and modify the mathematical description of such an effect, and to compare more in detail our results with the experimental data for non-resonant tunneling of photons. Finally, as a second example, the tunneling phase-time is calculated, and compared with the available experimental results, in the case of the quantum-mechanical tunneling of neutrons through two barrier-filters at the *resonance* energy of the set-up.

1. **Introduction.**

In this paper we study the behavior of the phase-time for tunneling through *two* successive potential barriers, completing our previous theoretical results[1] [which were confined to the rather interesting analysis of non-resonant tunneling through two opaque barriers] by extending them to the general case of resonant and non-resonant tunneling of a (non-relativistic) particle through a generic double-barrier --without forgetting about the analogous case of "photon tunneling".

We shall compare our predictions with the experimental data, which are scarce in the case of quantum-mechanical tunneling, but are numerous in the analogous, corresponding case of "classical barriers" (i.e., of tunneling photons or evanescent waves).

As to quantum-mechanical (resonant) tunneling, we shall compare our results with ref.[2], where experimental data on the time spent by cold neutrons when tunneling through a so-called neutron interference filter have been presented. The structure of the filter used therein consists in two rectangular barriers separated by a rectangular potential well; and the directly measured quantity has been the precession-phase shift for neutrons during their interaction with the neutron-filter. On the basis of their measurements, the authors of ref.[2] have actually obtained the tunneling Larmor time for neutrons in the region of the relevant resonance, corresponding to a quasi-bound state of the neutrons in the potential well between the barriers. In refs.[2-4] it was shown that, however, for a sufficiently large region of magnetic field and for quasi-monochromatic neutrons, the Larmor-time coincides with the phase-time.

As to the "classical tunneling", we shall be able to compare our results both with resonant and non-resonant tunneling experiments, in particular through two successive classical barriers.

## 2. The resonant tunneling case: Conditions.

Let us consider the tunneling of a particle with effective mass $m$ and kinetic energy $E$ through *two equal rectangular barriers* with width $a$ and height $U_0$, separated by the distance $L$ (cf. Fig.1). The amplitude of the transmitted wave in this case is defined by the formula [5]

$$A_T(k) = \exp(-2ika)/D(k) \qquad (1)$$

where

$$D(k) = \cosh^2(qa) + \frac{1}{4}\sinh^2(qa)\left[\sigma^2 \cos(2kL) - \delta^2\right] +$$
$$+ i\sinh(qa)\left[\delta \cosh(qa) + \frac{1}{4}\sigma^2 \sinh(qa)\sin(2kL)\right], \qquad (2)$$

$k = \sqrt{2mE}\,2\pi/h$, $q = \sqrt{2m(U_0 - E)}\,2\pi/h$, $\delta = (q^2 - k^2)/kq$, and $\sigma = (k^2 + q^2)/kq$. The dimensionless constants $\delta$ and $\sigma$ are connected by the relation

$$\sigma^2 = \delta^2 + 4. \qquad (3)$$

At a resonance, the double barrier becomes totally transparent[5] (i.e., without any reflections), so that at the resonance it is characterized by the condition

$$|A_T(k)|^2 = 1. \qquad (4)$$

It is easy to see (cf. Eq.(2), and also ref.[5]) that the values of the wave number $k$ for which condition (4) is satisfied can be found from the equation

$$\cot(kl) = -\frac{1}{2}\delta \tan(qa). \qquad (5)$$

We can show that from eq.(5) it is possible to find out also the values of parameters $a$, $m$ and $U_0$, at the resonance. Indeed, from eqs.(1) and (4) it follows that at a resonance it is

$$|D(k)|^2 = 1. \qquad (6)$$

On introducing the functions

$$u = \cosh^2(qa) - \frac{1}{4}\delta^2 \sinh^2(qa), \quad v = \delta\cosh(qa)\sinh(qa), \quad w = \frac{1}{4}\sigma^2 \sinh^2(qa) \qquad (7)$$

and using eq.(3), one can infer that these functions are connected by the relation

$$u^2 + v^2 = (1+w)^2. \qquad (8)$$

Then, by using functions (7), the denominator of eq.(2) can be written in the form

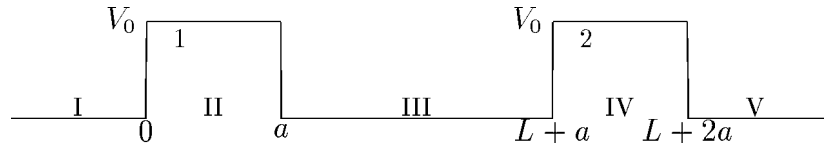

Figure 1: Scheme of the tunneling process through *two* successive (opaque) potential barriers. We examine in this paper the total Tunneling Times corresponding to this set-up, both in the case of resonant and non-resonant tunneling, and show their very remarkable peculiarities.

$$D = u + w\cos(2kL) + i[v + w\sin(2kL)]. \tag{9}$$

It follows from eqs.(9) and (8) that

$$|D|^2 = 1 + 2w[1 + w + u\cos(2kL) + v\sin(2kL)], \tag{10}$$

and condition (6) gets transformed into

$$1 + w + u\cos(2kL) + v\sin(2kL) = 0. \tag{11}$$

Taking eq.(3) into account, we can write the term $1+w$ of eq. (11) in the form

$$1 + w = \cosh^2(qa) + \frac{1}{4}\delta^2 \sinh^2(qa),$$

and this (last) equation can be easily transformed, afterwards, into eq.(5). There are no other solutions to eq. (11). Hence, eq.(5), as well as eq.(4) or eq.(6), is a general resonance condition, and it can be used for finding the value of any one of the parameters $a$, $k$, $L$, $m$ and $U_0$ at the resonance, once the values of all the other parameters are known. For instance, in the experimental paper [2], the values of $a$, $L$ and $U_0$ were known, and it was measured the "resonance value" of the neutron energy; in such a case, eq. (5) allows us calculating the value of the effective neutron mass and using it, then, for the evaluation of the tunneling phase-time.

3. **Behavior of the Transmission Coefficient at a resonance.**

In the region of a resonance, if we limit ourselves to the first two terms of the expansion of eq.(2) into a series of powers of $E - E_r$ (quantity $E_r$ being the resonance value of energy $E$), we obtain

$$D(k) = D_r + C_r(E - E_r), \tag{12}$$

where $D_r = D(k_r)$, $k_r = \sqrt{2mE_r}\, 2\pi/h$ ,

$$C_r = \frac{4\pi^2 m}{h^2 k} D'_r, \tag{13}$$

and the index *prime* defines the derivative with respect to $k$. We can rewrite eq.(12) in the form

$$D(k) = C_r(E - E_r + D_r C_r^* / |C_r|^2). \tag{14}$$

It follows from eq.(5) that at the resonance it is

$$\cos(2k_r L) = -u_r/(1 + w_r), \quad \sin(2k_r L) = -v_r/(1 + w_r), \tag{15}$$

where $u_r$, $v_r$ and $w_r$ are the values at $E = E_r$ of the functions $u$, $v$ and $w$, respectively. On inserting eq.(15) into eq.(9), we obtain

$$D(k_r) = (u_r + iv_r)/(1 + w_r). \tag{16}$$

Differentiating eq.(9) with respect to $k$, and inserting the result into eq.(15), we get

$$D'(k_r) = [u'_r - (u_r w'_r - 2Lv_r w_r)/(1 + w_r)] + i[v'_r - (v_r w'_r - 2Lu_r w_r)/(1 + w_r)]. \tag{17}$$

Then, on using eqs.(8), (13), (16), (17) and relation

$$uu' + vv' = (1 + w)w', \tag{18}$$

which follows from eq.(8), one finds that at the resonance

$$D_r C_r^* = i \frac{m}{\eta^2 k_r} \left[ \frac{u'_r v_r - u_r v'_r}{1 + w_r} + 2Lw_r \right], \tag{19}$$

where [by having recourse to eqs.(8),(13),(17) and (18)] quantity $\eta$ represents here $h/2\pi$ and

$$|C_r|^2 = \left(\frac{m}{\eta^2 k_r}\right)^2 \left[(u'_r)^2 + (v'_r)^2 - (w'_r)^2 + 4l\frac{u'_r v_r - u_r v'_r}{1 + w_r} w_r + 4L^2 w_r^2\right]. \tag{20}$$

From eqs.(8) and (18) one gets

$$u'^2 + v'^2 - w'^2 = (u'^2 + v'^2)^2/(1+w)^2, \tag{21}$$

while, from eqs.(20) and (21), one obtains

$$|C_r|^2 = \left(\frac{4\pi^2 m}{h^2 k_r}\right)^2 \left(\frac{u'_r v_r - u_r v'_r}{1 + w_r} + 2Lw_r\right). \tag{22}$$

By differentiating the functions (7) with respect to $k$, one can see that

$$u'_r v_r - u_r v'_r = \frac{1}{q_r}(1 + w_r)/[\delta_r k_r a + 2q_r Lw_r + \sigma_r^2 \cosh(q_r a)\sinh(q_r a)], \tag{23}$$

where $q_r$, $\delta_r$ and $\sigma_r$ are the values of $q$, $\delta$ and $\sigma$ at $E = E_r$. On inserting eq.(23) into eqs.(19) and (22), we can then rewrite eq.(14) in the noteworthy form

$$D(k) = C_r(E - E_r + i\beta), \tag{24}$$

where

$$\beta = \frac{h^2 k_r q_r}{4\pi^2 m} [\delta_r k_r a + 2q_r Lw_r + \sigma_r^2 \cosh(q_r a)\sinh(q_r a)]^{-1}, \tag{25}$$

so that eqs.(14), (19), (22) and (24) yield

$$|C_r|^2 = 1/\beta^2. \tag{26}$$

Finally, by inserting eq.(24) into eq.(1) and taking account of eq.(26), we obtain that near a resonance it holds in general

$$|A_T(k)|^2 = \frac{\beta^2}{(E-E_r)^2 + \beta^2} \; , \qquad (27)$$

which corresponds to nothing but a Breit and Wigner formula! In other words, one verifies that the Breit-Wigner's formula has a general validity for our (one dimensional) tunnelling near a resonance[*].

### 4. Phase time for (resonant and/or non-resonant) tunneling: A general formula.

Let us start from the relation

$$\tau = (h/2\pi) d \arg\{A_T \exp[ik(2a+L)]\}/dE \; , \qquad (28)$$

obtained by us, in ref.[1], for the tunneling phase-time in the case of *two* successive barriers (and which is valid whenever the constant-phase approximation can be adopted, i.e., when the waves traveling back and forth between the two barriers can be regarded as forming two packets only, one traveling forward and the other backwards).
By using eqs.(1) and (9), we can rewrite eq.(28) in the form

$$\tau = \frac{2\pi m}{hk}\left[L - (D_1 D_2' - D_2 D_1')/|D|^2\right] \qquad (29)$$

where

$$D_1 = u + w\cos(2kL), \quad D_2 = v + w\sin(2kL), \qquad (30)$$

[the index *prime* denoting again the derivative with respect to $k$]. On inserting eq.(30) into eq.(29) and using eq.(8), we get *in general* for the total tunneling phase-time the remarkable formula

$$\tau = \frac{2\pi m}{hk}\frac{P}{|D|^2} \; , \qquad (31)$$

where

$$P = (1+2w)L + u'v - uv' + (u'w - uw')\sin(2kL) + (vw' - v'w)\cos(2kL) \; . \qquad (32)$$

It should be noticed that eq.(31) holds in general *for any (resonant and/or non-resonant) tunneling*

---

[*] Let us recall that the width of the resonances decreases for increasing $L$.

*time* through two barriers. From eqs.(31),(32), (6) and (19), it follows that that the tunneling phase-time at a resonance $\tau_r$ is given by the expression

$$\tau_r = \frac{2\pi m}{hkq}\left[\sigma^2 \cosh(qa)\sinh(qa) + \delta ka + (1+2w)qL\right]. \tag{33}$$

### 5. Energy behavior of the tunneling phase-time near a resonance.

It follows from eqs.(1) and (28) that

$$\tau = (h/2\pi)\mathrm{d}\arg[\exp(ikL)/D]/\mathrm{d}E. \tag{34}$$

Inserting eq.(24) into eq.(34), we get

$$\tau = (h/2\pi)\mathrm{d}\arg[\exp(ikL)/C_r(E - E_r + i\beta)]/\mathrm{d}E,$$

so that, near a resonance, the behavior of the tunneling phase-time in terms of the energy is represented by the interesting formula

$$\tau \cong \frac{2\pi m}{hk}L + (h/2\pi)\frac{\beta}{(E-E_r)^2 + \beta^2}, \tag{35}$$

holding for any resonant tunneling through the two barriers. The first term represents the time associated with the particle free-flight over the distance *L* between the two barriers; while the second term is the time delay caused by the quasi-bound state assumed by the particle in such an intermediate region.

### 6. About the dependence of the tunneling phase-time on the lenth of the set-up, far from the resonances.

It was shown in our paper, ref.[1], that, when *a* is large enough (and *q* not too small) so that $qa \to \infty$, one happens to get for non-resonant tunneling the rather interesting relation

$$\tau \to 4\pi m/hkq, \tag{36}$$

so that, while depending on the energy of the tunneling particle, does depend neither on a nor on *L*. This result not only confirmed the so-called ``Hartman effect''[4] for the two opaque barriers --i.e., the independence of $\tau$ from the opaque barrier widths--, but also extended such an effect by implying the considered total tunneling time to be independent even on the distance between the barriers. Actually, eq.(36) shows that τ is given solely by the time spent inside the first one of the two opaque barriers, with no contribution coming, in particular, from the intermediate region between the barriers. [To avoid confusions, let us warn the reader that the distance $L$ between the barriers in refs.[1] was called $L - a$.] One might be tempted to regard this as a further evidence of the fact that quantum systems seem to behave as non-local, but it has actually a more general meaning, it being associated with the properties of any waves (and, in fact, something very similar is known[3,6] to happen also, e.g., with electromagnetic waves). It is important to stress once more that the previous result holds, however, for non-resonant (nr) tunneling [that is, far from the resonances that arise in the intermediate region due to interference between forward and backward travelling waves] and under the mentioned hypotheses only. However, even if striking, such a result has been theoretically confirmed and generalized (in terms of "superoscillations") by Aharonov et al. in ref.[7], and experimentally verified by Longhi et al.[8], as well in the pioneering experiments by Nimtz et al.[9]. For a more complete picture, see ref.[10].

When releasing the above condition, we have found in this paper a more complicate expression, given by formulae (31), (32) and (10) above. Anyway, from eqs.(1), (2), (7), (10), (28), (31) and (32) it follows that, for opaque barriers, i.e., when $qa >> 1$, it holds far from the resonances the important formula

$$\tau \approx \frac{4\pi m}{hkq} + 8\pi \frac{m}{hk} L \exp(-2qa) \left[ \frac{\sigma^2}{4} + \left(1 + \frac{\delta^2}{4}\right) \cos(2kL) + \delta \sin(2kL) \right]^{-1}, \qquad (37)$$

which can explain, as we are going to see, the experimental results recently found by Longhi et al.[8] and by Nimtz[11].

Namely, when $qa$ increases, the second term in eq.(37) decreases as $\exp(-2qa)$; while, at the limit when $qa \to \infty$, eq.(37) goes into eq.(36), a result first obtained, as already mentioned, in our ref.[1]. We would like now to underline that, in quantum experiments, the tunneling-time of a non-relativistic particle is expected to be practically measurable for large $qa$ values only (but not too large, of course, to avoid that the tunneled particles are too few). Therefore, in order to be able to reproduce theoretically any experimental results, it seems to be necessary studying the behavior of the transmission coefficient for large $qa$. Indeed, from eqs.(1) and (27) it follows that

$$|A_T|^2 \approx 32 \frac{1}{\sigma^2} \exp(-4qa) \left[ \frac{1}{4}\sigma^2 + \left(1 - \frac{1}{4}\delta^2\right) \cos(2kL) + \delta \sin(2kL) \right]^{-1}, \qquad (38)$$

which shows that, with increasing $qa$, the transmission coefficient square $|A_T(k)|^2$ decreases as $\exp(-4qa)$, that is, even more quickly than the second term in the r.h.s. of eq.(37). Consequently, from the experimental point of view, *no tunneling can take place for those values of $qa$ which make negligible the mentioned second term* in eq.(37). Moreover, for the values of $qa$ for which the tunneling probability is experimentally significant, the second term in the r.h.s. of eq.(37), which is slowly exponentially decreasing with $qa$, results to depend very weakly on the distance $L$ between the barriers: more precisely, it will depend almost linearly (but very slightly) on $L$, with even slighter oscillations. This explains the experimental fact that τ has been actually observed to increase (very slowly, almost linearly, and probably with very small oscillations) on $L$: cf., e.g., Fig.5 in ref.[8] and ref.[11]. Of course, our previous results do not hold only for particles but –as well-known[6]– also for photons.

Let us add the incidental comment that the last observations of ours may agree with the interpretation, of the prediction associated with eq.(37), forwarded by Aharonov et al.[7] within the super-weak interaction approach. Indeed, it is evident from eqs.(37)-(38) that the "generalized Hartman effect" (GHE) tends to appear for a *small* subclass of the tunneling particles; and the more the subclass gets small, the more the GHE holds and becomes impressive.

### 7. Brief comparison with the experimental data on resonant neutron tunneling.

In the experiments reported in ref.[2], a neutron flux was used with wave length $\lambda = 20{,}1 \overset{\circ}{A}$ and spectrum half-width $\Delta\lambda/\lambda \cong 4{,}8\%$ The width of the angular distribution of the neutron flux was 3.2 mrad, while the measurements were performed for sliding (gliding) angles. The neutrons in ref.[2] tunneled through two equal rectangular barriers with width $a = 300 A°$ and height of about 230 neV, separated by a potential well with width $L = 195 \, A°$ (and almost zero value of the potential). For such a set-up, a single resonance level exists, with energy $E_r \approx 127$ neV and half-width 4 neV. The measured neutron time delay at the resonance was $(2.17 \pm 0.2) \cdot 10^{-7}$ s (whilst in the regions far from the resonance it was $1.9 \cdot 10^{-8}$ s, due to non-tunneling neutrons).

Our calculations on the base of formula (5) show that, for the given experimental values of the parameters $a$, $L$, $U_0$ and for an effective neutron mass $m$ equal to the mass of the free neutron $m_0$, the resonance value of the energy is equal to 123 neV. The exact experimental resonance value of the neutron energy $E_r = 127$ neV can be obtained from eq. (5), by contast, for $m = 0.926883 \, m_0$. Adopting such a such value of the neutron effective mass, the value of the tunneling time, calculated with the help of the formulae (33) and (25), results to be $2.36 \cdot 10^{-7}$ s. The difference between the experimental and the calculated tunneling time at the resonance may be explained, on one hand, by the experimental neutron non-monochromaticity (and angular distribution) as well as by the energy resolution of the detectors, and, on the other hand, by the approximate character of the tunneling phase-time itself. Since the experimental set-up inevitably causes an averaging of the resonance tunneling time over a certain energy interval, for a good comparison we have to average the calculated tunneling phase-time near the resonance over the interval $[E_r - \beta, E_r + \beta]$, thus obtaining the value $\bar{\tau} = 2.4 \cdot 10^{-7}$ s, which is again near to the experimental value. Of course, for more rigorous calculations, more realistic (not quasi-monochromatic) wave packets are to be used [4,5], while the hypotheses underling the adoption of the tunneling phase-time should be probably revised.

### 8. Some conclusions.

General expressions have been firstly obtained for the transmission coefficient and tunneling phase-time in the case of resonant and non-resonant particle tunneling through two equal rectangular barriers; and comparisons with experimental data have been performed, which confirm that the simple use of Wigner's phase-time for evaluating the tunneling times leads to satisfactory results in all the considered cases. In particular, a comparison with the data[2] for cold-neutron tunneling through a neutron filter has been performed; and a good agreement with (and explanation for) such experimental data has been obtained. Moreover, a semi-quantitative explanation for the non-resonant tunneling-time behavior observed in recent optical two-barrier experiments [8,11] has been forwarded.

**Acknowledgements**

For useful, stimulating discussions, the authors are grateful to F.Bassani, G.Degli Antoni, H.E.Hernàndez-Figueroa, J.Leòn, G.Nimtz, R.Mignani, D.Mugnai, M.Pernici, V.Petrillo, B.Reznik, R.Riva, G.Salesi, and particularly to R.Bonifacio, S.De Leo, S.Esposito, P.Rotelli, M.Villa, B.N.Zakhariev and M.Zamboni-Rached.